\documentclass[showpacs]{revtex4}
\usepackage{amsmath}
\usepackage{latexsym}
\usepackage{epsfig}

\newcommand{\lta}{\;
  \raise0.3ex\hbox{$<$\kern-0.75em\raise-1.1ex\hbox{$\sim$
  }}\;\hskip-2pt }
\newcommand{\gta}{\;
  \raise0.3ex\hbox{$>$\kern-0.75em\raise-1.1ex\hbox{$\sim$
  }}\;\hskip-2pt }

\begin{document}

\title{The growth of human settlements during the Neolithic, clustering and
food crisis}
\author{ Sergei Fedotov, David Moss, Daniel Campos}

\address{School of Mathematics, The University of Manchester, Manchester, UK }

\date{\today}

\begin{abstract}
We present a stochastic two-population model that describes the
migration and growth of semi-sedentary foragers and sedentary
farmers along a river valley during the Neolithic transition. The
main idea of this paper is that random migration and transition from
sedentary to foraging way of life and backward is strongly coupled
with the local crop production and the associated degradation of
land. We derive a non-linear integral equation for the population
density coupled with the equations for the density of soil nutrients
and crop production. Our model provides an explanation for the
formation of human settlements along a river valley. The numerical
results show that the individual farmers have a tendency for
aggregation and clustering. We show that the large-scale pattern is
a transient phenomenon which eventually disappears due to land
degradation.
\end{abstract}

\maketitle
\affiliation{School of Mathematics, University of Manchester, Manchester M60 1QD, UK}

\section{Introduction}

The wave of colonization by migrating farmers and establishment of farming
communities in Europe between $5000$ and $3500$ BC is currently a topic of great
interest in prehistoric archaeology, linguistics and anthropology \cite%
{Ch,Pr}. Ammerman and Cavalli-Sforza developed a model for the expansion of
farming as a demic diffusion which spread into Europe in the form of wave of
advance \cite{a,a2}. Using the radiocarbon dates, they found that farmers
spread at an average rate of about one kilometer a year. Interest in simulation
and spatial modelling of spread of agriculture has been growing rapidly in
the last decade, especially in the physics community
\cite{fort1,VR,fort2,fort3}%
. One of the main reasons for this is that the geographical spread
of population can be effectively described by the classical
Fisher-KPP equation and its various generalizations \cite{E}. These
models have attracted considerable interest in physics and biology,
because of the huge number of potential  applications. Fort and
Mendez have applied a time-delayed theory for the Neolithic
transition \cite{fort1} which involves a hyperbolic correction to
the Fisher-KPP equation. The transition from hunter-gathering to
farming did not happen in a uniform way, and that is why Davison et
al. have taken into account both advection and spatial variation in
diffusivity and carrying capacity, in the framework of the
Fisher-KPP equation \cite{Dav}. Aoki and Shida \cite{Aoki} have
studied the spread of farmers into an area occupied by
hunter-gatherers in terms of a system of reaction-diffusion
equations. The main objective of those works was to reproduce the
observed rate at which agricultural expansion took place in Europe.
Despite the interest in establishment of farming communities in
Europe, there remains no published material on the spatial structure
formed behind the wave of advance. The main challenge of our work
presented below is to set up a model which provides an explanation
to not only the propagation of farming but also the formation of
human settlements. We develop a model that demonstrates the tendency
of the distribution of population to form clusters,  and
furthermore, that this large-scale pattern in population evolution
is a transient phenomenon, which disappears due to degradation of
land in the form of an extinction wave. Our specific motivation is
the successive migration of settlements along parallel river valleys
in the Tripolye-Cucuteni system, which has been thoroughly
investigated and documented, e.g. \cite{dol2}. This system can be
considered as being one-dimensional, and so is particularly amenable
to numerical investigation.

\section{A two-population model}

\subsection{A two-population model for semi-sedentary foragers and sedentary
farmers}

In our model the population consists of semi-sedentary foragers and
sedentary farmers who share the same territories. The semi-sedentary
foragers are the population of individuals randomly moving from
place to place along a river valley and searching for food and other
resources. An implicit consequence of this behaviour is the
foundation of new settlements (large localized values of population
density), and an interchange between farming and foraging
populations. On the contrary the sedentary farmers are individuals
who do not migrate. They live in small villages scattered near
cultivated land in the major river valleys. Their main activities
are the cultivation of soil and crop
production. In this paper we are interested in the total population density $%
n(x,t)$ at location $x$ along the river at time $t.$ We define this density
as $n=n_{1}+n_{2}$, where $n_{1}$ is the density of semi-sedentary foragers
and $n_{2}$ is the density of sedentary farmers.

We assume that there is not a strict distinction between foraging
and sedentary lifestyles. There are always random transitions from a
sedentary to a foraging way of life, and {\it vice versa}; these
transitions depend strongly on the local food supply. This is one of
the main features of our random walk model. Regarding the movement
of semi-sedentary foragers, they do not jump from place to place
completely randomly. Unlike Brownian particles in physics, the
migration of people cannot be explained by a standard diffusion law,
in which the flux is proportional to the gradient of number density
of individuals. To describe the random migration of semi-sedentary
foragers and random transitions from one lifestyle to another we
adopt a biased random walk whose statistical characteristics depend
on the local food supply. In our model, the probability of a random
migration event making a jump $z$ in the time interval $t$ to
$t+\Delta t,$ is $\lambda \Delta t.$ The probability of transition
from foraging lifestyle to the farming is $\alpha _{1}\Delta t.$ The
probability of the conversion of farmers to semi-sedentary foragers
is $\alpha _{2}\Delta t.$ Thus we introduce a new variable, the
local crop production per individual per year $q(x,t),$ so that the
frequency of jumps $\lambda $ and transition rates $\alpha _{1}$ and
$\alpha _{2}$ depend on the crop production:
\begin{equation*}
\lambda =\lambda (q),\quad \alpha _{i}=\alpha _{i}(q)\quad i=1,2.
\end{equation*}%
It is natural to assume that the frequency $\alpha _{2}(q)$ is a
decreasing function of $q.$ That is, when sedentary farmers are not
able to produce enough food to sustain their population, some of
them start to migrate from their neighborhood at rate $\lambda
(q(x,t))$.

\subsection{Balance equations for population densities}

We now set up the balance equation for the density of semi-sedentary
foragers, $n_{1}(x,t).$ According to our random walk model, the density of
foragers $n_{1}$ at location $x$ at time $t+\Delta t$ can be written as
follows
\begin{eqnarray}
n_{1}(x,t+\Delta t) &=&\left( 1-\lambda (q(x,t))\Delta t-\alpha
_{1}(q(x,t))\Delta t\right) n_{1}(x,t)+  \notag \\
&&\int \lambda (q(x-z,t))\Delta tn_{1}(x-z,t)\rho (z)dz+\alpha
_{2}(q(x,t))\Delta tn_{2}(x,t).  \label{n1}
\end{eqnarray}%
This equation is a conservation law for foragers. The first term on
the right hand side represents those foragers who stay at location
$x$ and do not move during time $\Delta t$ and do not become
sedentary farmers. The
second term gives the number of foragers who arrive at $x$ during time $%
\Delta t$ from different places $x-z$, where the jump distance $z$
is distributed by dispersal kernel $\rho \left( z\right) .$ The last
term is the number of sedentary farmers who become the
semi-sedentary foragers during the time $\Delta t.$

The sedentary farmers do not migrate, and their density $n_{2}(x,t)$ obeys the
balance equation involving logistic growth and lifestyle transitions:
\begin{eqnarray}
n_{2}(x,t+\Delta t) &=&\left( 1-\alpha _{2}(q(x,t))\Delta t\right)
n_{2}(x,t)+  \notag \\
&&rn_{2}(x,t)\left( 1-\frac{n(x,t)}{K\left( q\right) }\right) \Delta
t+\alpha _{1}(q(x,t))\Delta tn_{1}(x,t),  \label{n2}
\end{eqnarray}%
where $r$ is the growth rate of the sedentary population. The
carrying capacity $K$ is in general, an
 increasing function of the local crop production $q$.
The last term gives the number of foragers who become the farmers during the
time $\Delta t.$ In the limit $\Delta t\rightarrow 0$, from (\ref{n1}) and (%
\ref{n2}) we obtain two differential equations for $n_{1}(x,t)$ and $%
n_{2}(x,t):$
\begin{equation}
\frac{\partial n_{1}}{\partial t}=\int \lambda (q(x-z,t))n_{1}(x-z,t)\rho
(z)dz-\lambda (q)n_{1}-\alpha _{1}(q)n_{1}+\alpha _{2}(q)n_{2}  \label{nnn1}
\end{equation}%
\begin{equation}
\frac{\partial n_{2}}{\partial t}=rn_{2}\left( 1-\frac{n}{K\left( q\right) }%
\right) +\alpha _{1}(q)n_{1}+\alpha _{2}(q)n_{2}.  \label{nnn2}
\end{equation}

In this paper we are interested in the evolution of population density $%
n(x,t)$ on characteristic time scales around $100-500$ years.
Therefore we can adopt a local equilibrium for the two populations,
describing the balance between the nomadic and sedentary ways of
life in proportions $p$ and $1-p.$ We write
\begin{equation}
n_{1}(x,t)=pn(x,t),\qquad n_{2}(x,t)=\left( 1-p\right) n(x,t),  \label{main}
\end{equation}%
where the proportion of foragers $p$ is the function of the crop production $%
q:$
\begin{equation}
p=p(q).  \label{main2}
\end{equation}%
In the asymptotic regime when \ $\alpha _{i}>>\lambda ,$ one can write the
dependence of $p$ on the transition rates $\alpha _{1}(q)$ and $\alpha
_{2}(q)$ as
\begin{equation}
p(q)=\frac{\alpha _{2}(q)}{\alpha _{1}(q)+\alpha _{2}(q)}.
\end{equation}%
The evolution equation for biased migration (\ref{nnn1}) and population
growth (\ref{nnn2}) can be rewritten as one balance equation for the total
population density$.$ By adding equations  (\ref{nnn1}) and (\ref{nnn2})
and using (\ref{main}) we obtain a single equation for $n(x,t)$
\begin{eqnarray}
\frac{\partial n}{\partial t} &=&\int p(q(x-z,t))\lambda
(q(x-z,t))n(x-z,t)\rho (z)dz  \notag \\
&&\ -p(q(x,t))\lambda (q(x,t))n(x,t)+(1-p(q))rn\left( 1-\frac{n}{K\left(
q\right) }\right) .  \label{3}
\end{eqnarray}%
Now we need to find an approximate form for the function $p(q).$ It
is natural to assume that for a large crop production $q,$ the
population mostly consists of sedentary farmers, that is, $p$ close
to zero, while for a low crop production the population mostly is
semi-sedentary foragers ($p$ is close to one). A simple
approximation for $p$ might be
\begin{equation*}
p(q(x,t))=\mathcal{H}\left( q(x,t)-q_{\min }\right) ,
\end{equation*}%
where $\mathcal{H}(x)$ is the Heaviside step function:
$\mathcal{H}(x)=0$ if $x>0;$ $\mathcal{H}(x)=1$ if $x\leq 0.$ Thus
farmers do not migrate if the crop supply is greater than a critical
value $q_{\min }$.  In our numerical simulations we use a piece-wise
approximation: a linearly decreasing function in the interval
between the minimum value $q_{\min }$ and the maximum value $q_{\max
}$
\begin{equation*}
p(q(x,t))=\left\{
\begin{array}{c}
p_{\rm max},\;q\leq q_{\min }, \\
-aq+b,\;q_{\min }<q<q_{\max }, \\
p_{\rm min},\;q\geq q_{\max }%
\end{array}%
\right. .
\end{equation*}%
Based on \cite{dol}, we take $q_{\min },q_{\max }=300,736$ kg per
person per year respectively. For the results discussed below we set
$p_{\rm max}=0.95$, $p_{\rm min}=0.05$ (see Fig. 1) but the exact
values are not crucial to our results.

The dispersal kernel $\rho (z)$ is assumed to have two cutoffs: a
minimum cutoff (foragers do not travel very small distances) and a
maximum cutoff (foragers do not migrate over very large distances in
a single jump). In numerical simulations we use minimum cutoff
$\Delta z$ of between $5$ and $20$ km. The maximum cutoff is more
difficult to estimate. However, since the typical length of the
river is around $500$ km, it seems reasonable to assume that the
maximum cutoff is around $50$ km (see Fig. 1).

\subsection{Equation for crop production}

We assume that the human population derives most of their food from
the cultivation of land. Thus we suggest the following formula for
the local food production $q(x,t)$
\begin{equation}
q(x,t)=\alpha \left( \frac{n(x,t)}{n_{0}+n(x,t)}\right) \left( 1-e^{-\beta
F(x,t)}\right) ,  \label{4}
\end{equation}%
where $F(x,t)$ is the density of soil nutrients, $\alpha $ is the
production rate coefficient, and $\beta $ is the parameter that
determines how the yield depends on the nutrients. This equation
describes how the rate of food supply $q$ increases due to the
increase in the population density $n$, and how the degradation of
land (the decrease of soil nutrients $F)$ leads to a decrease of
food production through the factor $1-e^{-\beta F(x,t)}.$ Note that
the factor $\frac{n(x,t)}{n_{0}+n(x,t)}$ describes a tendency toward
group solidarity that increases the efficiency of food production.
We know from existing data that the yield, i.e. the production of
food per unit area, can be up to $0.0736$ kg/m$^{2}$year
\cite{dol,har}.We estimate that the area cultivated by people is
approximately $10^{4}$ m$^{2}$ per person. Thus, we have an estimate
for $\alpha$ of  $736$ kg per person per year. In general, it is
difficult to model how the production of food depends on the
nutrients, since this relation is strongly dependent on the
environmental conditions. One of the existing models is given by the
Mitscherlich-Baule yield response \cite{Mit} which takes into
account productivity losses due to soil erosion. Following this
model,  in \cite{hop} the factor $1-e^{-\beta F(x,t)} $ has been
proposed, and this is what we have introduced into our model. The
value $\beta \simeq 890$ m$^{2}$kg$^{-1}$ is found for the corn
cultivation. However, other studies \cite{ove} seem to suggest lower
values ($\beta \simeq 65$ m$^{2} $kg$^{-1}$).

\begin{figure}[tbp]
\begin{tabular}{ll}
(a)\includegraphics[width=0.40\hsize]{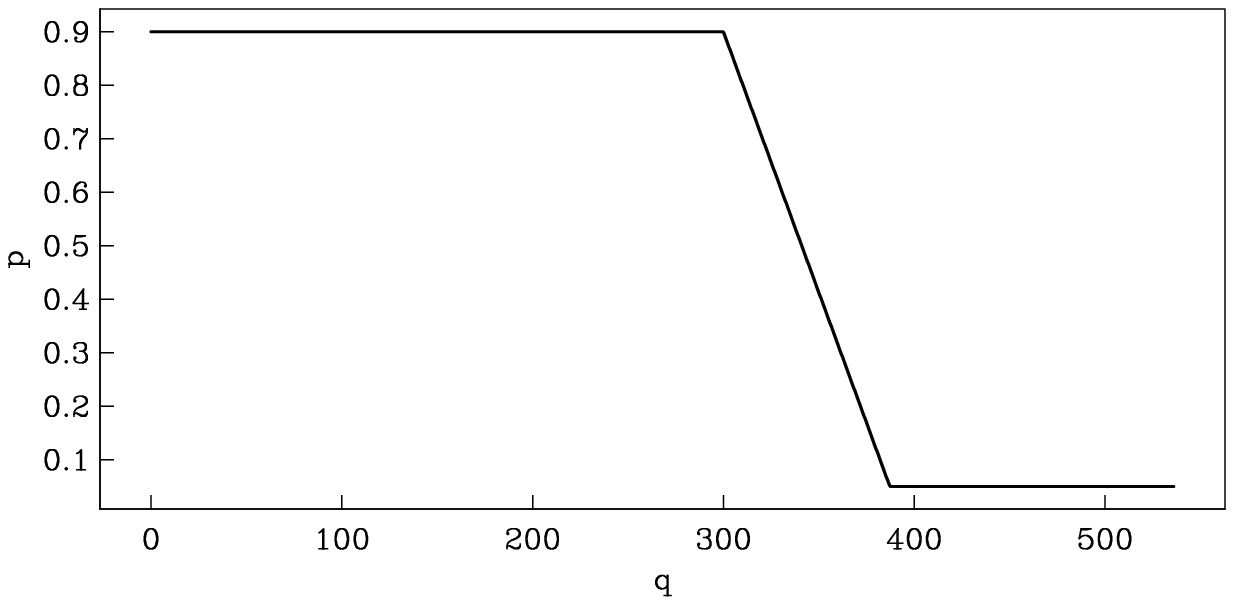}&
(b)\includegraphics[width=0.40\hsize]{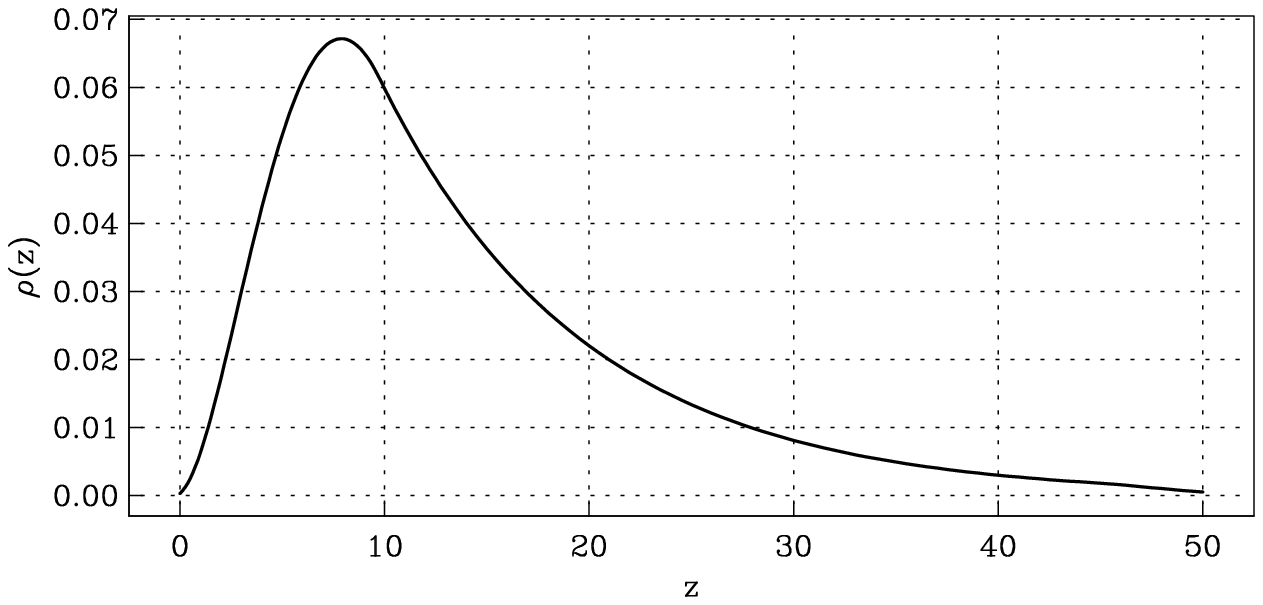}
\end{tabular}%
\caption{(a) The function $q(p)$ and (b) $\rho(z)$ for $\Delta z=10$
km. } \label{profiles}
\end{figure}

\subsection{The land degradation equation}

We adopt the following model for nutrient depletion and the corresponding
land degradation. We assume that equation for the density of soil nutrients $%
F(x,t)$ has the form
\begin{equation}
\frac{\partial F(x,t)}{\partial t}=\xi _{1}-\xi _{2}F(x,t)-\gamma
q(x,t)n(x,t)  \label{5}
\end{equation}%
where $\xi _{1}$ is the rate at which soil nutrients regenerate naturally, $%
\xi _{2}$ is the rate of nutrient depletion due to environmental
reasons (erosion, flooding, etc.), $\gamma $ is the rate at which
nutrients are depleted due to the harvests. Here we assume that
natural nutrient depletion is much slower than depletion due to
human activity. In order to estimate $\gamma $, we can use the data
from \cite{har}. This data corresponds to prehistoric agriculture in
Hawaii; however, similar values are cited for Sub-Saharan
agroecosystems and there is no data available (as far as we know)
for other regions. So using that study we estimate $4\cdot 10^{-4}$
kg$_{P}$/m$^{2}$yr, where kg$_{P}$ denotes kilograms of phosphorus
in the soil. This value, together with the data above of $0.0736$
kg/m$^{2}$yr lead us to assume that the rate of nutrient depletion
$\gamma =5.43\cdot 10^{-3}$ kg$_{P}$/kg . However, other studies
mention that the total concentration of nutrients in crops can be up
to 3\% \cite{ove}. It would suggest that the value $\gamma =0.03$ kg$_{P}$%
/kg can be taken as a maximum threshold. Of course, other nutrients
are also important, but phosphorous can be regarded as a proxy for
them. Soil regeneration is a very complex process and depends on
many environmental parameters. Nevertheless, in some agroeconomics
models regeneration is modeled as a constant \cite{mcco} as we have
assumed in our model.

\section{Numerical results}

Numerical simulations of the non-linear integral equation (\ref{3})
together with (\ref{4}) and (\ref{5}) reveal a very interesting
dynamical behaviour: the emergence of large-scale patterns in
population density. This phenomenon can be interpreted as the
formation of human settlements along a river valley. Thus our model
provides an explanation for formation of settlements as a dynamical
phenomenon. The individual farmers have a tendency for aggregation
and clustering as a result of non-linear random migration. Moreover
our model describes not just a population clustering but subsequent
decay of these clusters (settlements) due to land degradation.

Our model has a rather large number of parameters, but fortunately
most of them can be estimated from existing sources, as indicated
above. For numerical
work we use as unit of time $1$ yr, measure densities in units of m$%
^{-2} $, and assume $\lambda =\mathrm{const}$. We take the values
$r=0.03$, $\alpha =736,$ $\beta =200,$ $K=10^{-4},$ $\xi
_{1}=10^{-4},$ $\xi _{2}=0,$ $q_{\min }=300,$ $q_{\max }=736$. Note
that for simplicity in this initial study we take $K$ to be a
constant, rather than a function of $q$. We explore
ranges $0.01\leq \gamma \leq 0.03$ yr$^{-1}$, $0.05\leq \lambda \leq 0.3$ yr$%
^{-1}$, $5\leq \Delta z\leq 20$ km. Our initial conditions were rather
arbitrary: $n=0.1K$ in $0\leq x\leq 15$ km, $n=0$ elsewhere, although from
experiments with other initial conditions with $n(x,0)$ non-zero in $x\;%
\raise0.3ex\hbox{$<$\kern-0.75em\raise-1.1ex\hbox{$\sim$
  }}\;\hskip-2pt100$ it appears that results are quite insensitive to
details of the initial state. Again for simplicity, we set the
initial value $F(x,0)$ to be uniform and equal to $F_0$. Our
`standard' choice was $F_0=0.01$. If the range of $x$ were to be
extended to negative values, the evolution of $n(x,t)$ would be
approximately symmetric about $x=0$.

Fig. 2 shows the population density $n(x,t)$ for $%
\lambda=0.1$, with $\Delta x=10$, $\gamma=0.01$, at intervals of 100
yr. Panel (a) shows the "arrival" of farmers at location $0<x<15$
along the river. After $300$ years, one can see clear evidence of
clustering of population ("settlements"). Panel (e) shows their
subsequent decay and reappearance of clusters behind the extinction
wave. The entire population decays over about 800 yr, depending to
some extent on the exact choice of parameters. A general feature of
our modelling is that most population clusters grow and decay {\it
in situ}, without significant movement.
We can make an order of magnitude estimate of the total population
of a cluster by taking the linear extent of a cluster as the
diameter of a circular settlement. This gives typical figures of
$O(10^4)$ individuals.

\begin{figure}[tbp]
\begin{tabular}{ll}
(a)\includegraphics[width=0.40\hsize]{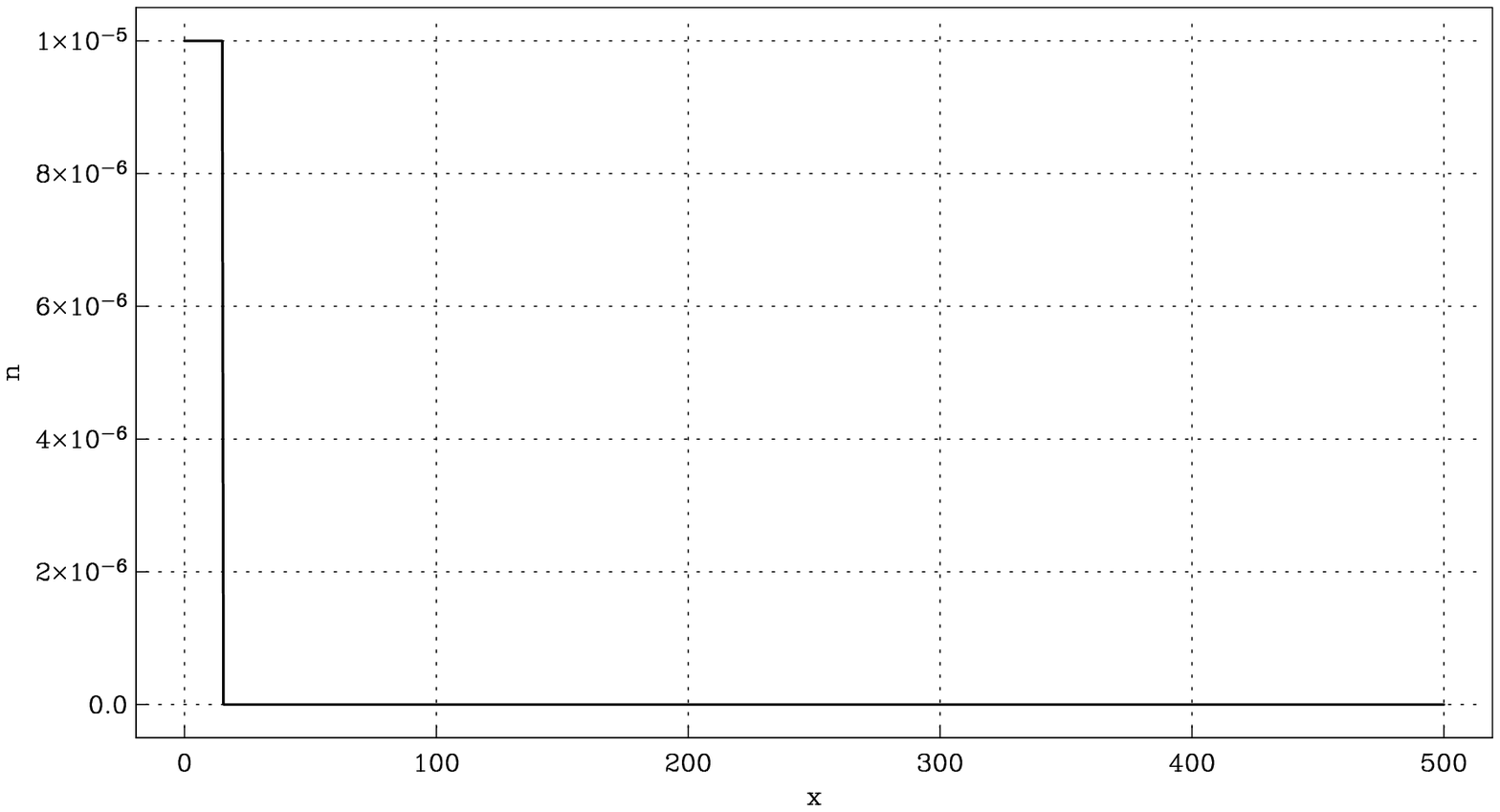} (d)%
\includegraphics[width=0.40\hsize]{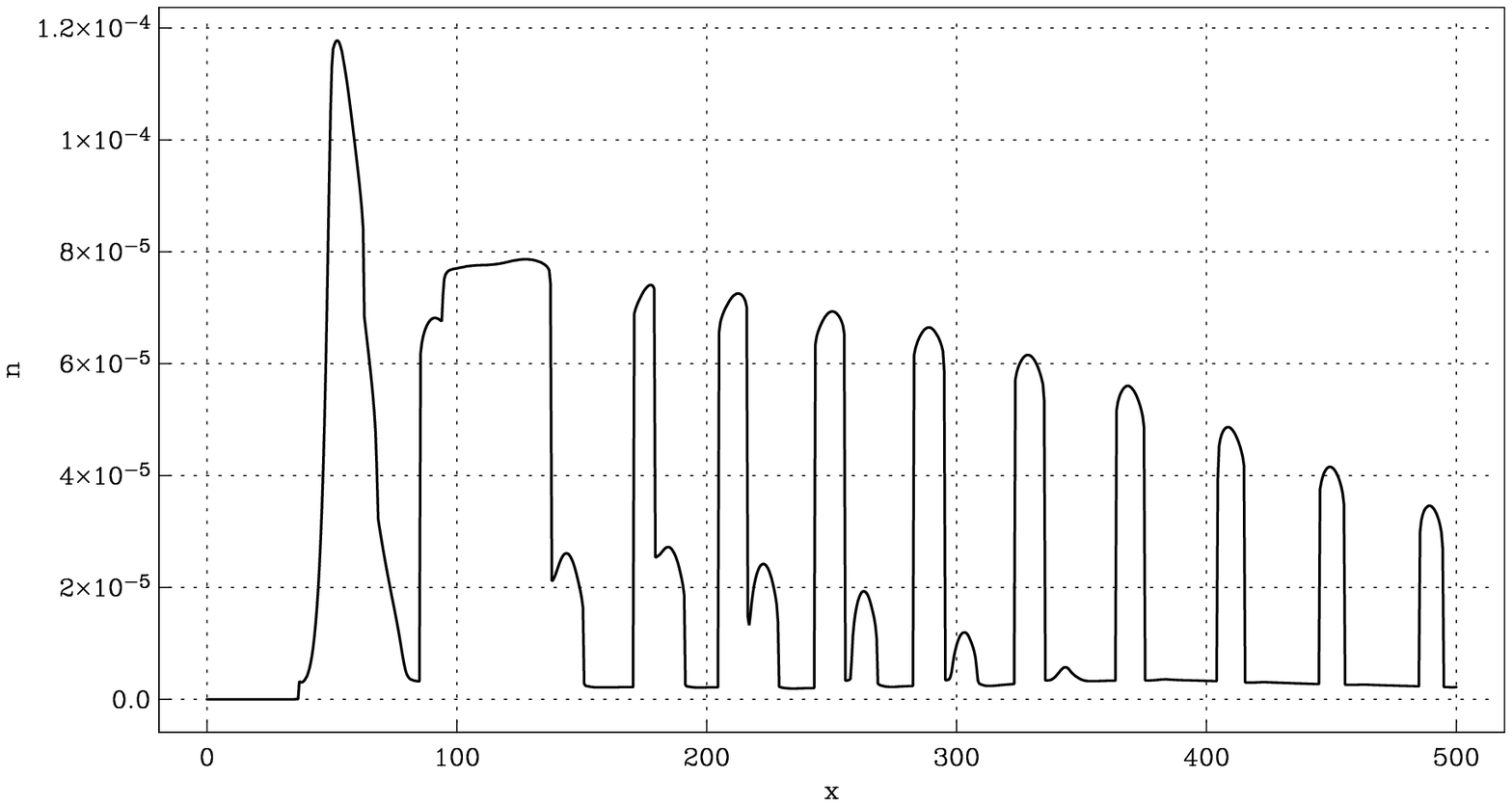} &  \\
(b)\includegraphics[width=0.40\hsize]{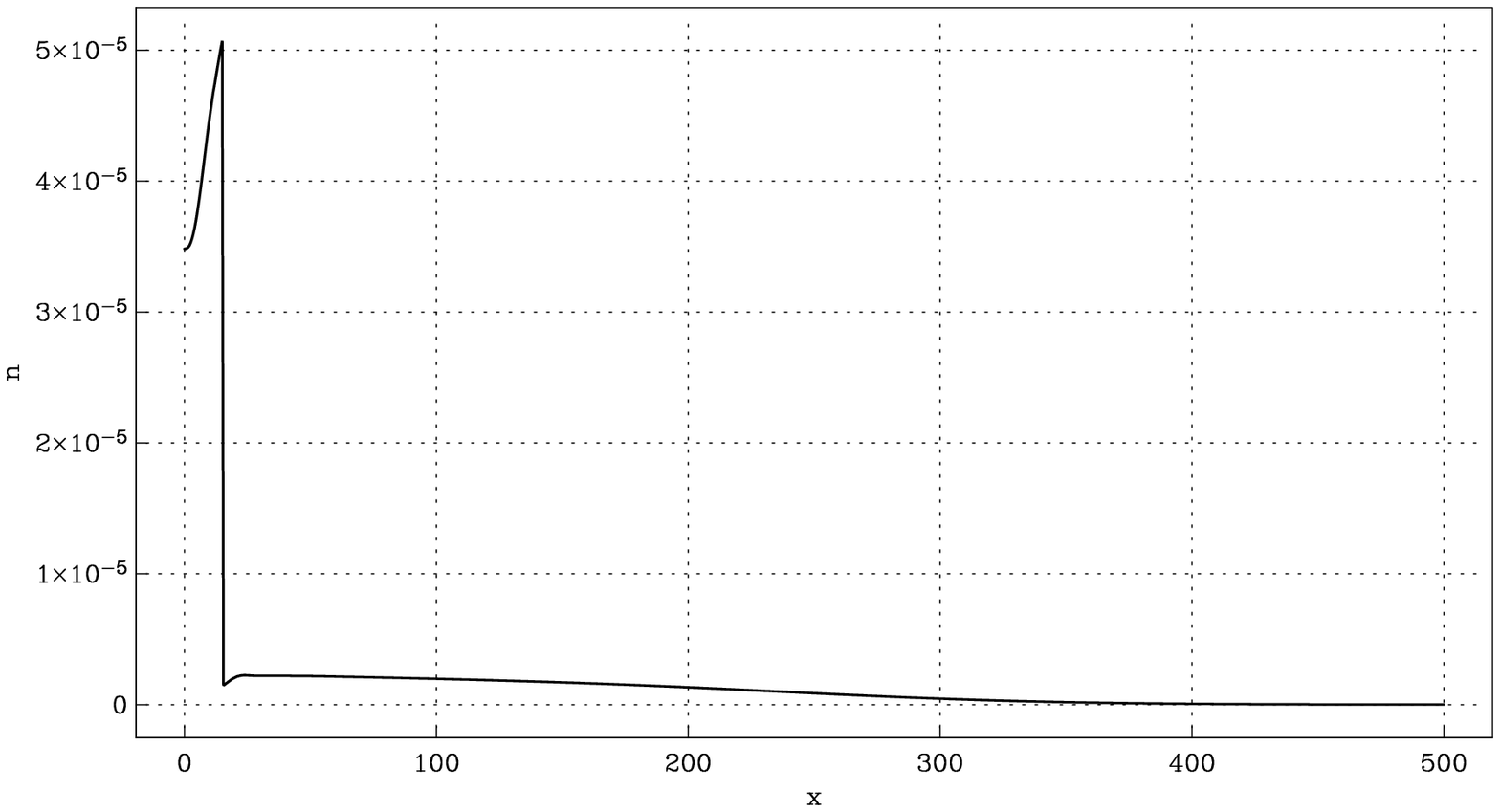} (e)%
\includegraphics[width=0.40\hsize]{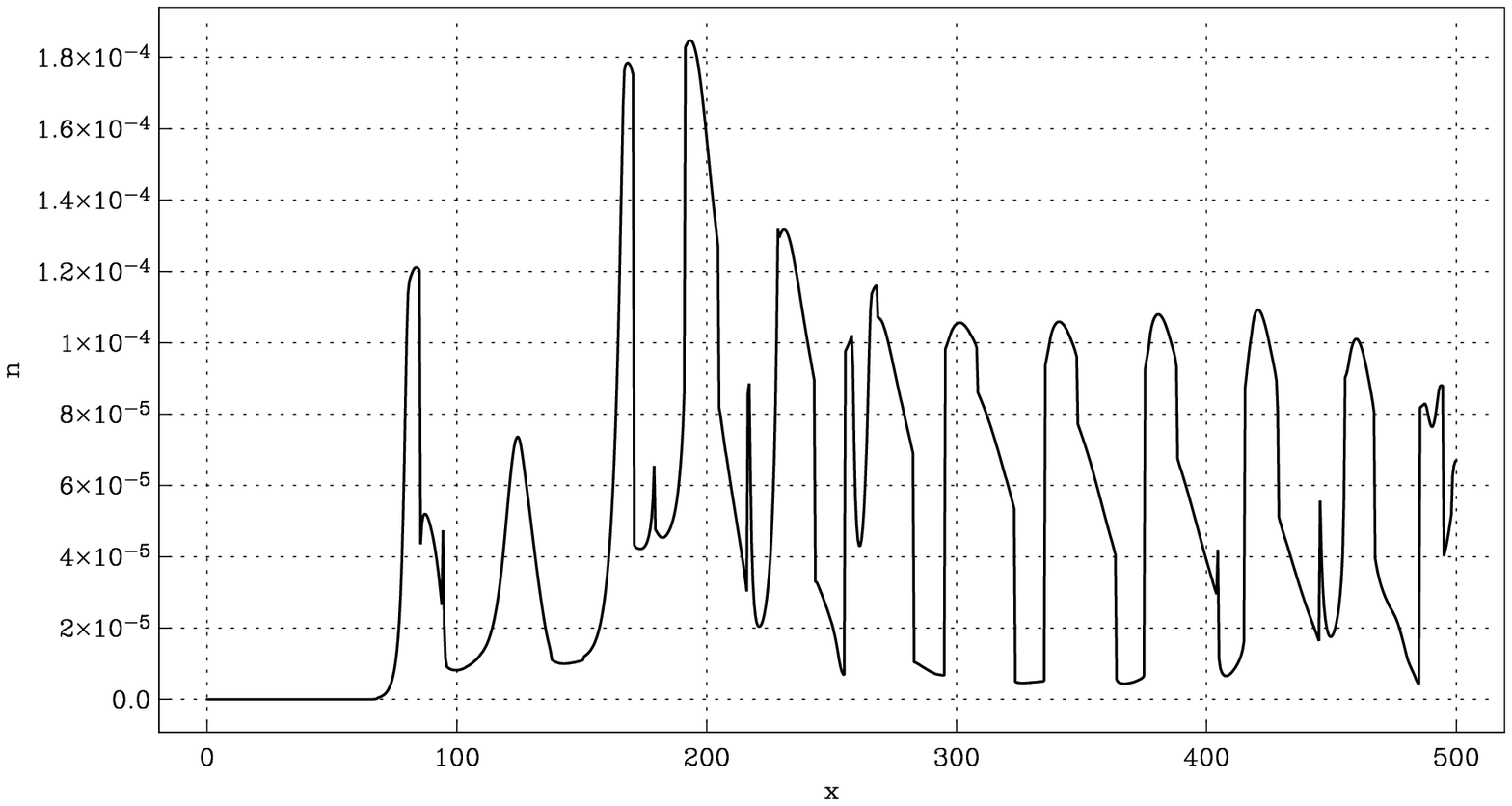} &  \\
(c)\includegraphics[width=0.40\hsize]{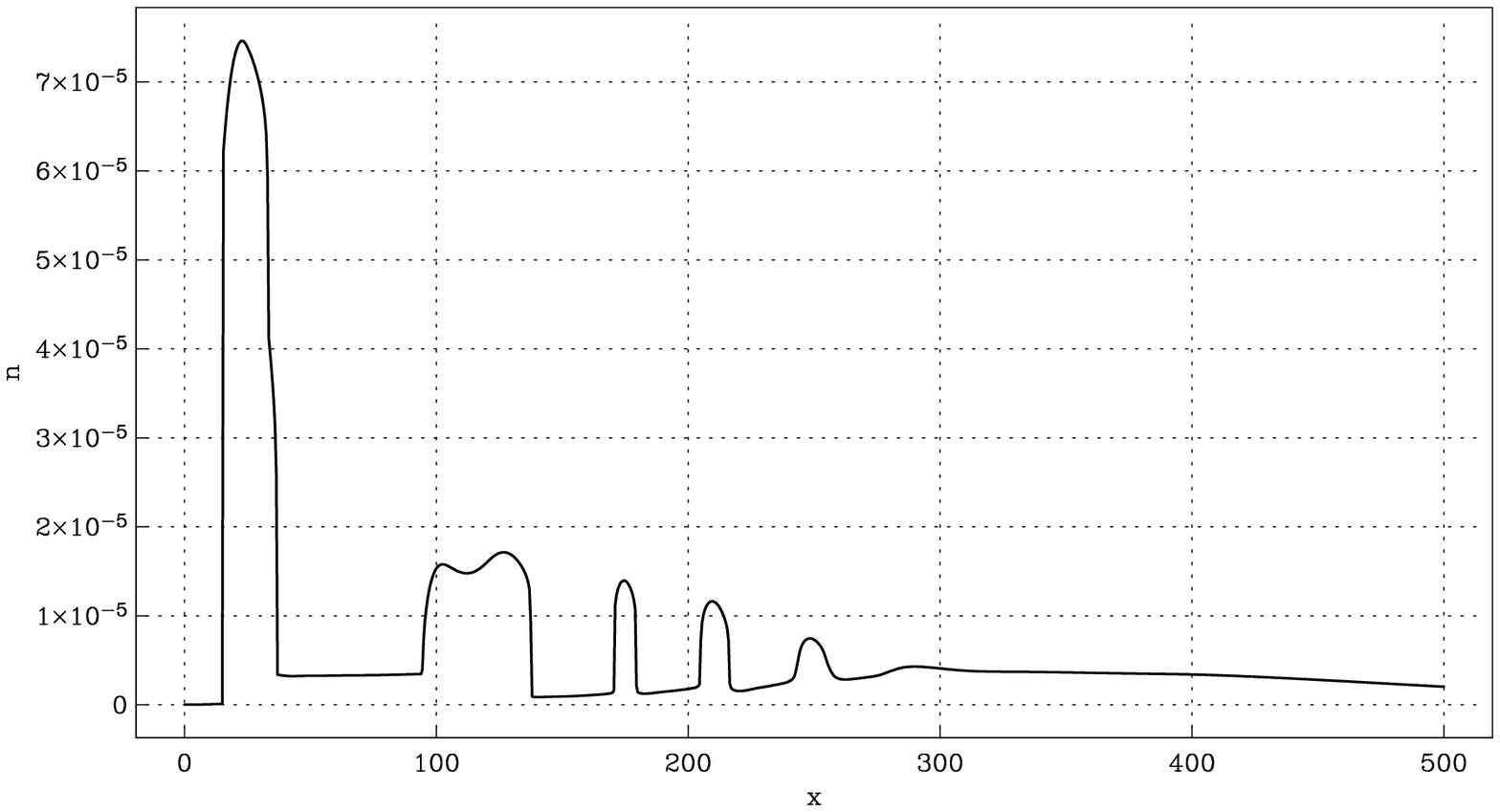} (f)%
\includegraphics[width=0.40\hsize]{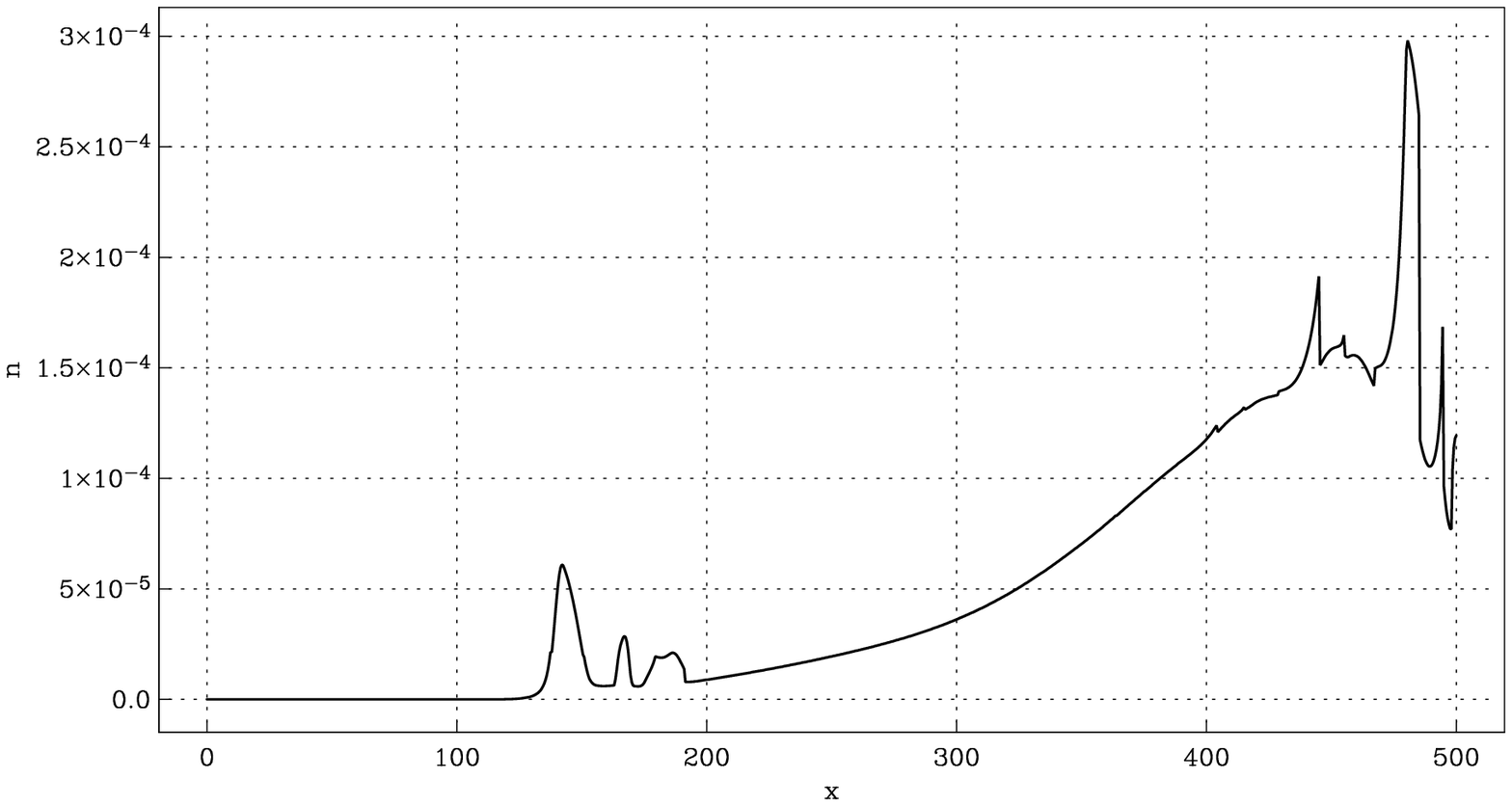} &  \\
\end{tabular}%
\caption{$n(x,t)$ for the model with $\protect\lambda=0.2$, $\protect\gamma%
=0.01$, $\Delta x=10$ discussed in the text. The panels (a) -- (f)
show successively the solution at times $0 - 500$ yr, at intervals
of $100$ yr. } \label{nlambda2}
\end{figure}


We also explored the dependence of the speed of population advance,
and the separation of the clusters, as a function of the parameters.
The speed of advance is rather ill-defined, but for simplicity we
monitored the movement of the main peak of the distribution.
(Arguably we could have, e.g., studied the position of the cluster
furthest from the origin.) We
found that the speed of advance depends approximately linearly on both $%
\lambda$ and $\gamma$, and to be insensitive to $\Delta x$. The separation
of clusters is independent of both $\lambda$ and $\gamma$ and depends
approximately linearly on $\Delta x$ ($= (1.5-2) \Delta x$). If $\lambda\; %
\raise0.3ex\hbox{$<$\kern-0.75em\raise-1.1ex\hbox{$\sim$
  }}\;\hskip-2pt 0.05$ or $\Delta x\; \raise0.3ex%
\hbox{$>$\kern-0.75em\raise-1.1ex\hbox{$\sim$
  }}\;\hskip-2pt 15$ the clustering phenomenon does not occur.
 While we have concentrated our efforts on studying variation among a
manageable subset of parameters while retaining fixed plausible
values for the others, we did also make a small study of the effects
of varying $F_0$ and $\xi_1$. Increasing the value of $F_0$, up to
$0.1$ from our canonical value of $0.01$, prolongs the timescale of
the population evolution, by a factor of up to about $2$. Values of
$F_0$ significantly smaller than $0.01$ do not give clustering for
typical choices of the other parameters. It means that for
non-fertile soils transition to farming and the emergence of
clusters are unlikely to occur.
For larger values of $\xi_1, \; \raise0.3ex%
\hbox{$>$\kern-0.75em\raise-1.1ex\hbox{$\sim$
  }}\;\hskip-2pt 10^{-4}$, the regeneration is so strong that, although
clustering occurs initially much as for $\xi_1=10^{-4}$, the final state is
a spatially uniform population. For smaller values of $\xi_1$, e.g. $%
\xi_1=10^{-5}$, we still see strong clustering, but the overall phenomenon
persists for a significantly shorter time, as the episodes of cluster
regeneration seen when $\xi_1=10^{-4}$ do not now occur.

\section{Discussion and conclusions}

We have explored a continuous model of linear (one-dimensional)
population migration and clustering. We took our inspiration and
guidance from the successive south to north migration of the
Tripolye-Cucuteni cultures along parallel river valleys \cite{dol2}.
We attempted only to illustrate the migration along a single valley.
The main result is that our model gives an explanation for the
formation of settlements as a dynamical phenomenon. The individuals
have a tendency for aggregation and clustering as a result of
non-linear random migration. Moreover our model describes subsequent
decay of these clusters (settlements) due to land degradation in the
form of an extinction wave. Clearly our model, being mathematically
continuous, cannot be expected to reproduce in detail the
essentially discrete phenomenon of the establishment, growth and
decay of major settlements and their satellites, as revealed by the
archaeological record -- our goal is far more modest. For reasonable
choices of parameters we obtain migration over distances of order
$500$ km in times of $500-1000$ yr. Without fine tuning, our
solutions exhibit distinct clustering of population, at intervals of
$10-30$ km. Plausibly these clusters are estimated to have $O(10^4)$
individuals. These estimates are all consistent with what is known
about the Tripolye-Cucuteni cultures \cite{dol}. Overall, we feel
that our quite naive modelling captures some of the essence of the
clustering seen in population migration, and points the way for more
sophisticated modelling. We can think of a number of significant
developments in the future, including taking $K=K(q)$ and allowing
for a latency effect, in which population in a cluster
(`settlement') has a reduced probability of moving, representing an
attachment to the investment of building a house and clearing land.
It is easy to allow for additional resources available from e.g.
forest or river, which are maybe harder to exhaust than the
fertility of the land and so can act as a reservoir of resource. It
would be interesting to apply our model to understand societal
collapses to which environmental problems as habitat destruction,
soil degradation and overpopulation contribute \cite{dia}.

\section*{Acknowledgment} The research is supported by the European Community's Sixth Framework
Programme, grant NEST-028192-FEPRE.



\begin{thebibliography}{99}
\bibitem{Ch} V. G. Childe, The dawn of European civilisation. (London:
Routledge and Kegan Paul; 1968).

\bibitem{Pr} T. D. price (Ed.), Europe's first farmers (Cambridge University
Press, Cambridge, 2000).

\bibitem{a} A. J. Ammerman, \ and L. L. Cavalli-Sforza, Man \textbf{6}, 674
(1971).

\bibitem{a2} A. J. Ammerman, \ and L. L. Cavalli-Sforza, (1984) The
Neolithic Transition and the Genetics of Populations in Europe (Princeton
Univ Press, Princeton, 1984).

\bibitem{fort1} J. Fort, and V. Mendez, Phys Rev Lett \textbf{82}, 867
(1999); Phys. Rev. E \textbf{60}, 5894 (1999).

\bibitem{VR} M. O. Vlad, and J. Ross, Phys. Rev. E \textbf{66}, 061908 (2002)

\bibitem{fort2} J. Fort , T. Pujol, L. L. Cavalli-Sforza. Camb Archaeol J.
\textbf{14}, 53 (2004).

\bibitem{fort3} J. Fort, J. P\'{e}rez-Losada, N. Isern. Phys Rev E \textbf{76%
} 031913 (2007).

\bibitem{E} U. Ebert and W. van Saarloos, Physica D \textbf{146}, 1 (2000);
Phys. Rep. \textbf{337}, 139 (2000).

\bibitem{Dav} K. Davison, P. Dolukhanov, G. R. Sarson, and A. Shukurov J
Arch Sci \textbf{33}, 641 (2006).

\bibitem{Aoki} K. Aoki, S. Mitsuo and S. Nanako, Theor. Population Biology,
\textbf{50}, 1 (1996).

\bibitem{dol2} M. Zvelebil and P. Dolukhanov, J. of World Prehistory \textbf{%
5}, 233 (1991).

\bibitem{dol} P. Dolukhanov (private communication)

\bibitem{har} A. S. Hartshorn, O. A. Chadwick, P. M. Vitousek and P. V.
Kirch. Proc. Natl. Acad. Sci. USA \textbf{103}, 11092 (2006).

\bibitem{Mit} H. D. Patterson, Biometrics, \textbf{25}, 159 (1969).

\bibitem{hop} J. W. Hopkins, R. Lal, K. D. Wiebe and L. G. Tweeten. Land
Degrad. Develop. \textbf{12}, 305 (2001).

\bibitem{ove} A. R. Overman, R. V. Scholtz III and F. G. Martin. Commun.
Soil Sci. Plan. \textbf{34}, 851 (2003).

\bibitem{mcco} K. E. McConnell. An Economic Model of Soil Conservation. Am.
J. Agric. Econ. \textbf{65}, 83 (1983).

\bibitem{dia} J. Diamond, Collapse: How Societies Choose to Fail or
Succeed (Penguin books, London, 2005).


\end{thebibliography}
\end{document}